\documentclass{article}
\usepackage{spconf,amsmath,graphicx}
\usepackage{float}
\usepackage{multirow}
\usepackage{multicol}
\usepackage{tablefootnote}
\usepackage{xcolor}
\usepackage{subcaption}
\usepackage{amssymb}

\title{Two-pass endpoint detection for speech recognition}
\name{
\begin{tabular}{@{}c@{}}
Anirudh Raju$^{*}$ \thanks{\begin{tabular}{@{}l@{}}
$^{*}$Equal contribution. \\
Correspondence to \{ranirudh,apkhare,deehe\}@amazon.com
\end{tabular}},
Aparna Khare$^{*}$,
Di He$^{*}$, 
Ilya Sklyar, 
Long Chen, 
Sam Alptekin,\\ 
Viet Anh Trinh, 
Zhe Zhang, 
Colin Vaz, 
Venkatesh Ravichandran, 
Roland Maas, 
Ariya Rastrow
\end{tabular}}
\address{Amazon Alexa AI, USA}
\begin{document}

\maketitle
\begin{abstract}
Endpoint (EP) detection is a key component of far-field speech recognition systems that assist the user through voice commands. The endpoint detector has to trade-off between accuracy and latency, since waiting longer reduces the cases of users being cut-off early. We propose a novel two-pass solution for endpointing, where the utterance endpoint detected from a first pass endpointer is verified by a 2nd-pass model termed EP Arbitrator. 
Our method improves the trade-off between early cut-offs and latency over a baseline endpointer, as tested on datasets including voice-assistant transactional queries, conversational speech, and the public SLURP corpus. We demonstrate that our method shows improvements regardless of the first-pass EP model used.
\end{abstract}


\section{Introduction}

Speech as an input modality for human-machine interaction has become popular in recent years. 
In early voice-interface systems, users would indicate the start and end of speech by pressing and releasing buttons. However, with the rise of far-field speech recognition systems, there is a heightened emphasis on automated solutions for detecting the end of user speech, known as automatic endpoint detection or endpointing (EP). Automatic endpoint detection must balance the need to endpoint quickly, reducing the user's perceived system response time \cite{shangguan2021dissecting}, and the need to accurately endpoint to avoid cutting off the user's speech prematurely. Endpointing is challenging due to factors such as background noise, overlapping speech, and the necessity to support a diverse range of requests, including brief queries, lengthy and complex ones, as well as interactive conversations where the speaker may conclude their query and engage in concurrent conversations with someone nearby \cite{9383516,lu2022endpoint,chang2022turn}. Furthermore, the system needs to avoid cutting the user off prematurely due to within-utterance pauses, disfluencies, hesitations, false starts, and repetitions that are commonly present in natural speech.

Traditional endpointing approaches have relied on voice-activity detection (VAD) based solutions. The endpointer triggers when a non-speech period of a certain duration is detected. VAD models typically use deep neural networks to predict speech/non-speech activity for each feature-frame of audio \cite{ryant2013speech,hughes2013recurrent,thomas2015improvements,8461921}. 
Recently, standalone endpoint detection models have been trained to explicitly detect the end of speech (EOS) frame \cite{shannon2017improved}.
The standalone model can integrate the tasks of VAD and EP in the same model by classifying each of them per-frame in a multi-task framework \cite{maas2018combining}. 

These systems can be further improved by using features from the speech recognition's decoder in addition to the acoustic signal. This has been shown to improve model robustness in distinguishing within-utterance pauses from end-of-utterance pauses \cite{maas2018combining, liu2015accurate}. Lexical features have also shown to provide improvements \cite{ferrer2003prosody,arsikere2014computationally}. In recent years, automatic speech recognition (ASR) systems have been moving towards an end-to-end (E2E) architecture with a single model that performs the functions of acoustic, lexical and language models. Such E2E systems outperform conventional ASR systems when trained on large-scale datasets \cite{8268937}. There have been works on integrating endpointing directly into the E2E model, considering endpointing an extension of the ASR task. These models compute likelihood of one additional end of speech token to trigger endpointing directly \cite{8683109,huang2022e2e}. RNNT-based E2E models can delay speech tokens, but introducing a loss during training minimizes this delay, especially crucial for timely EOS token emission for endpointing \cite{yu21fastemit}. 
 There is also work introducing accurate time alignment information to the training of end of speech token to increase endpointing accuracy \cite{9054715,9383606}.
State-of-the-art systems usually combine decoding cues with acoustic-only detectors to achieve best endpointing performance \cite{8683109}. The acoustic and decoder cues are usually combined to allow each individual signal to trigger endpointing alone \cite{8683109}.

In this paper, we propose a novel approach to improve existing classes of endpointing systems, by adding a second-pass verification model we term the EP arbitrator. The EP arbitrator model gates the initial endpointing decision. An utterance is endpointed only when a candidate endpoint from the first-pass EP successfully passes through the EP arbitrator model. Unlike the streaming EP detector in the first-pass, the EP arbitrator utilizes segment-level encodings of the acoustics and recognition output available up to the candidate endpoint decision frame. Solutions for second-pass rescoring of recognition output have been widely studied in the context of real-time and low latency speech recognition systems \cite{sainath2019two}. However, running a second-pass for endpointing has not been considered before. \\ 
We show that this novel system comprising a second-pass EP arbitrator model: (1) provides improvements in early cut-offs where the user hesitates, or pauses longer after a semantically incomplete query (2) provides an improved trade-off between early EP rate and latency across various latency operating points for the system (3) generalizes to provide improvements in early EP rate on out-of-domain conversational data, without explicit training on conversational speech (4) provides improvements in WER and early EP rate for different baseline EP systems, either standalone or combined with an end-to-end speech recognizer. Finally, we set up a baseline EP system on the publicly available SLURP corpus \cite{bastianelli-etal-2020-slurp}, and show improvements in EP performance with the proposed method.

\section{Problem Definition}
We focus on the problem of endpoint detection for far-field voice assistants.  Endpoint detection is the task of determining the speech frame at which the user query ends. More formally, let the query be represented by an input speech feature frame sequence $\textbf{x} = (x_{1}, ..., x_{T})$, where $x_{t} \in \mathbb{R}^{|X|\times1}$. The true frame $x_{EOS}$ where the user query ends is termed end-of-speech (EOS). In practice, the ground truth EOS frame is determined by force aligning the input speech with the transcription. Endpointing latency is defined as the time (in milliseconds) between the ground truth EOS frame and the endpoint determined by an endpointer model. 

\section{Methods}
\vspace{-1mm}

We consider a real-time and low-latency speech recognition system for voice-based applications which require automatic endpointing to detect when the user has finished speaking, and to process the query and respond back. First, we describe  methods to model an endpointing system as a standalone detector or jointly with the speech recognizer in Sections \ref{sec:method-ep-standalone} and \ref{sec:method-ep-e2e}, following prior work. In Section \ref{sec:method-ep-arbitrator}, we introduce our proposed method, EP Arbitrator, as a second-pass EP module that can augment any existing EP detector.

\vspace{-2mm}
\subsection{Standalone EP detectors}
\vspace{-2mm}

\label{sec:method-ep-standalone}
One approach to build a standalone EP model involves using an acoustic-based binary classifier. This classifier is trained to classify each speech frame as either an end-of-speech (EOS) frame or otherwise, based on the current and prior speech frames. For each feature frame, it predicts a binary class probability denoted as $P(EOS | x_{1:t})$. A frame level cross-entropy loss is used to train the model. The frame-level labels are derived through forced-alignment of the speech with the transcription, which aids in identifying the EOS frame. Frames before the EOS frame are labeled as non-endpoints (negative label), and frames including and after the EOS are considered endpoints (positive label). At decoding time, the model output is gated on a decision threshold $P(EOS|x_{1:t})>T_{EP}$ to detect an endpoint. Typically, this is modeled with a streaming recurrent neural network architecture that consumes audio features $\textbf{x}$. 
\noindent A variant of this approach that still relies on acoustic-based signals, is to employ a multi-task model architecture that simultaneously models VAD and endpoint detection. The VAD predictions from this architecture can be leveraged to enhance EP detection performance by imposing guardrails, such as setting a minimum or maximum pause duration \cite{maas2018combining} for detecting endpoints.

\vspace{-2mm}

\subsection{End-to-end EP detectors}
\label{sec:method-ep-e2e}
\vspace{-2mm}

\label{sec:method-ep-e2e}
The EP decisions can be integrated directly into an E2E speech recognition system such as a recurrent neural network transducer (RNNT). The EP decision is detected at the frame where the EOS symbol is emitted by the speech recognizer \cite{8683109}. This method does not require forced alignment information during training, and can be trained directly using RNNT loss and speech transcriptions. This natively integrates both acoustic and decoder information to make the EP decision.

\begin{figure}
    \centering
    \includegraphics[width=8cm]{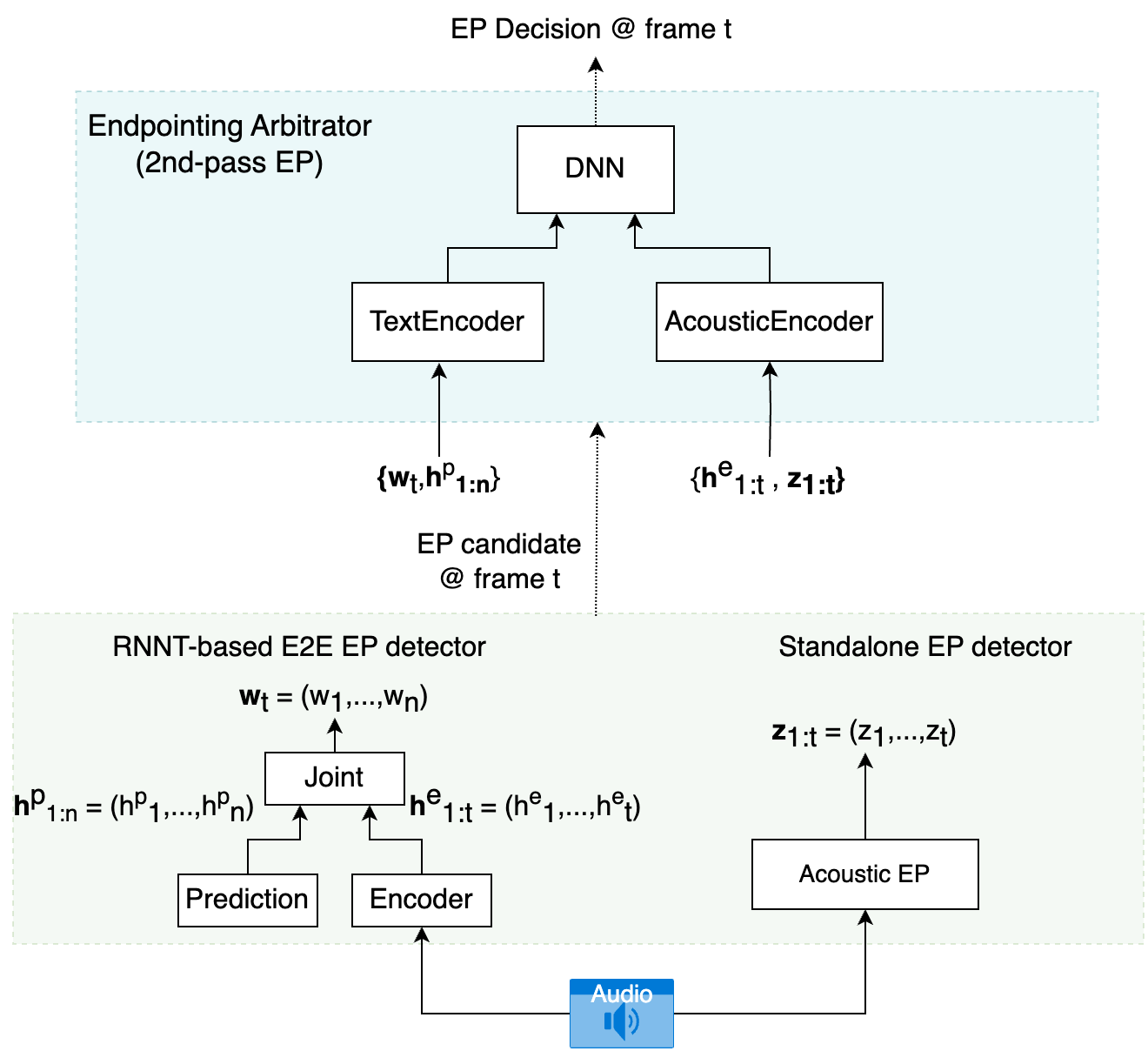}
    \caption{A candidate endpointing decision triggered at frame $t$ from the first-pass EP detector that is verified by the EP Arbitrator that consumes segment-level features of the acoustics and the recognition output}
    \label{fig:arbitrator-detailed}
    \vspace{-2mm}
\end{figure}

\vspace{-2mm}

\subsection{EP Arbitrator: second-pass detector}

\label{sec:method-ep-arbitrator}
The EP Arbitrator can be integrated on top of any type of EP detector as a 2nd-pass verification model. It verifies the first-pass candidate endpointing decision to emit the final system endpoint. Figure \ref{fig:arbitrator-detailed} shows the schematic of a system with EP Arbitrator integrated with an existing EP detection system. The existing EP detection system, termed the first-pass endpointing system, includes either standalone EP detectors, E2E detectors, or a combination of both. Any safety guardrails that are typically based on a pause-duration upper-bound, are exempt from gating by the EP arbitrator, to ensure the guardrail behavior is retained.

The EP Arbitrator triggers on candidate frame $t$ generated by the first-pass EP. It makes a segment-level endpointing decision by integrating cues from the 1st-pass endpointer, the speech recognizer, and the audio signal. In Figure \ref{fig:arbitrator-detailed}, a scenario is illustrated with a candidate endpoint at time $t$, that is emitted from either an RNNT EP or a standalone acoustic-based EP. The EP Arbitrator can access the input audio features and the intermediate hidden representations from either detector. Specifically, at frame $t$, the RNNT model has encoder representations $\textbf{h}^{e}_{1:t}=(h^{e}_{1},...h^{e}_{t})$ where $h^{e}_{t} \in \mathbb{R}^{|E|\times1}$, a partial one-best hypothesis token sequence of length $n$, $\textbf{w}_{t} = (w_{1}...w_{n})$ where $w_{i} \in \mathcal{W}$, $\mathcal{W}$ is the set of sentence-piece tokens used to train the RNNT model, and corresponding hidden representations from the prediction network $\textbf{h}^{p}_{1:n} = (h^{p}_{1},...h^{p}_{n})$ where  $h^{p}_{i} \in \mathbb{R}^{|P|\times1}$. The acoustic endpointer has hidden representations $\textbf{z}_{1:t} = (z_{1},...z_{t})$ where $z_{t} \in \mathbb{R}^{|Z|\times1}$ from an intermediate layer that is useful for making an EP decision. The EP Arbitrator uses a mix of these features to compute a segment-level EP decision posterior $P_{arb}(EOS | \textbf{x}_{1:t}$) at the candidate frame $t$ as follows:

\vspace{-4mm}
\begin{equation}
\vspace{-3mm}  
P_{arb}(EOS | \textbf{x}_{1:t}) = \text{ArbitratorModel}(\textbf{h}^{e}_{1:t}, \textbf{z}_{1:t}, \textbf{h}^{p}_{1:n}, \textbf{w}_{t})
\end{equation}
\\
This ArbitratorModel() leverages segment-level information of the acoustics from either the RNNT encoder ($\textbf{h}^{e}_{1:t}$) or the standalone acoustic-endpointer ($\textbf{z}_{1:t}$), or both. Further, it can leverage decoder information in the form of the 1-best recognition hypothesis ($\textbf{w}_{t}$), or the prediction network embeddings from the RNNT ($\textbf{h}^{p}_{1:n}$), or a combination of both. The ArbitratorModel() is modeled as a neural network.

The neural network architecture for EP arbitrator must be designed for low latency, to ensure that it does not delay the EP decision. Although the EP Arbitrator does not need to run on every frame, it may run on successive back-to-back frames if they satisfy the first-pass EP detection criteria. To minimize the latency impact, we decouple the ArbitratorModel into an AcousticEncoder that summarizes the acoustic information into an embedding, and a TextEncoder to summarize the information from the one-best recognition into another embedding. This decoupling allows for caching the 1-best text embedding for the EP Arbitrator posterior computation at all time frames  where the recognition output $\textbf{w}_{t}$ is unchanged.
\vspace{-2mm}
%
\begin{align}
\label{eqn:acoustic-encoder} 
\mathbf{v_{t}} &= \text{AcousticEncoder}(\mathbf{h}^{e}_{1:t}) \\
\label{eqn:text-encoder} 
\mathbf{e_{t}} &= \text{TextEncoder}(\mathbf{w}_{t}) \\
\label{eqn:arbitrator-softmax}
P_{arb}(EOS | \mathbf{x}_{1:t}) &= \text{softmax}(\text{DNN}(\text{concat}(\mathbf{v_{t}}, \mathbf{e_{t}})))
\end{align}

\vspace{-1mm}
The AcousticEncoder can accept input acoustic embeddings either from the RNNT, the acoustic EP, or both. As an example, Eqn \ref{eqn:acoustic-encoder} uses the RNNT encoder embeddings $\mathbf{h}^{e}_{1:t}$ only. By using these RNNT embeddings as input, it can leverage the ever-increasing scale of RNNT models to continue to improve accuracy. The AcousticEncoder summarizes the embeddings over the segment to $\textbf{v}_{t}$ with architectures such as choosing the input embedding of the current frame $t$, pooling the embeddings of the past frames and further processing with a deep neural network (DNN), or using a transformer-based encoder with pooling on the past frames. The TextEncoder summarizes the 1-best recognition output into a single embedding $\textbf{e}_t$. It first converts the token sequence $\textbf{w}_{t}$ into an embedding sequence using either a separate embedding, or reuses the embeddings sequence that correspond to $\textbf{w}_{t}$ from the RNNT prediction network at time $t$. This sequence is then consolidated into a single vector $\textbf{e}_{t}$ with architectures such as a DNN followed by pooling, a recurrent network, or a transformer-based model. Use-cases with short utterances can leverage bidirectional encodings to encode the segment-level recognition information without significant impact on latency. The acoustic and text embeddings are concatenated and followed with softmax classification to compute $P_{arb}(EOS | \mathbf{x}_{1:t})$ as in Eqn \ref{eqn:arbitrator-softmax}. The final EP decision is computed with a detection threshold $T_{arb}$. It should be noted that we do not use $\textbf{z}_{1:t}$ and $\textbf{h}^{p}_{1:n}$ in any of our experiments and leave their inclusion to future work. The exact model architecture used in our experiments is described in section \ref{sec:exp-setup-arbitrator}.

\vspace{-4mm}
\section{Experimental Setup}
\subsection{Datasets}
\vspace{-2mm}

\label{sec:datasets}
We report experimental results  both on the publicly available SLURP corpus, and on in-house voice-assistant datasets.
\vspace{-5mm}
\subsubsection{SLURP dataset}
There are no standardized publicly available datasets for the endpointing task. Prior works on endpointing have presented results on non-public datasets making the techniques difficult to reproduce \cite{maas2018combining,huang2022e2e,9054715}. In this work, we present endpointing results on the SLURP dataset \cite{bastianelli-etal-2020-slurp}, a publicly available speech corpus, commonly used to report speech recognition and natural language understanding metrics. It consists of 40.2 hours of training, 6.9 hours of development and 10.3 hours of evaluation data. 

The original SLURP utterances are pre-segmented with a very short duration of non-speech audio following the end of user's speech. To evaluate endpointing latency on this dataset, we would need to consider cases where the endpointer triggers after end of the original audio segments. To consider all such cases for evaluation, we simulate the audio following the end of user speech by padding all utterances in the SLURP dataset with 2 seconds of digital silence. Although this simulation isn't ideal and doesn't consider background noise following the user's query, it enables an apples-to-apples comparison of endpointing metrics across both the baseline and the proposed method.  We report experimental results on this dataset to show the efficacy of the method on a publicly available speech corpus, and help the method be reproducible for follow-up work in the speech community.

\subsubsection{Voice-assistant dataset}
The voice assistant dataset comprises de-identified US English far-field speech data directed to voice assistants. This dataset is representative of real-world conditions that include utterances with background noise.

\noindent \textbf{Training data}:
The training set includes 1000-hrs of human transcribed speech consisting of transactional queries that were directed to voice-assistants. This dataset was used to train both the baseline standalone EP system, and the EP Arbitrator, for all experiments. The ASR system used in all experiments is trained on a 100K-hrs training corpus of voice-assistant queries.

\noindent\textbf{Evaluation data}: We evaluate on three in-house test sets:
\vspace{-1mm}
\begin{enumerate}
\item
\noindent \textit{\textbf{Transactional queries dataset}}: This includes 50-hrs of development and test data consisting of queries directed to voice-assistants. The data was collected to ensure that the user's full query was captured with no cut-offs.

    \item 
\noindent \textit{\textbf{Transactional queries partial test set}}: The partial dataset includes 91-hrs of test data, collected with the endpointer described in section \ref{subsec:baseline}. These utterances were filtered to only include the instances where the user was cut-off before completing their sentence. The partial utterances were identified with a system that leverages models capable of determining semantic incompleteness and hesitation, from the recognition output and user feedback \cite{kim2021deciding}. 

\item
\noindent \textit{\textbf{Conversational test set}}: This includes de-identified conversations of users speaking with a voice-assistant socialbot \cite{ram2017conversational}. We leverage this test set to demonstrate generalization of our approach to more free-form, less transactional interactions with a smart assistant. To ensure higher semantic variability, this dataset is filtered to only include utterances with more than 10 words.

\end{enumerate}

\subsection{ASR system details}
\label{sec:model_details}

Acoustic features used for the RNNT ASR model are 64-dim log-mel filterbank energies computed over a 25ms window with 10ms shifts, and are stacked and downsampled to a 30ms frame rate. The acoustic features used by the standalone endpointing detector in the first-pass are the same. The ASR model has 148M parameters; it has an LSTM-based 8x1280 encoder, a 2x1280 prediction network, and a single-layer joint network \cite{graves2013speech}. The acoustic encoder of the  EP arbitrator model consumes the embeddings from the audio encoder of this ASR model. This model was trained on a large training corpus of over 100K-hrs. For the voice-assistant experiments, the trained ASR model is not updated further. \\
For the SLURP experiments, we fine-tune the ASR model described above on the train partition of the SLURP data for 10,000 steps, with a batch size of 32 and a learning rate of $3\text{e-}5$ with the Adam optimizer. The fine-tuning of pre-trained models to improve ASR performance on SLURP data is in line with prior work \cite{sunder2023fine}.\\
This ASR model, which is also used as E2E EP, is trained with fast-emit \cite{yu21fastemit} to minimize label emission delay and the fast-emit parameter $\lambda$ is tuned on a held-out development set to improve latency without degrading WER. Additionally, the model uses beam-search decoding which allows for the 1-best hypothesis to change after an initial EOS is emitted, as the model keeps decoding till the final endpoint is triggered.
\vspace{-3mm}
\subsection{Endpointing system details}

\subsubsection{Baseline EP system}

\label{subsec:baseline}
The baseline endpointing system uses a 1st-pass EP detector, that comprises two endpoint detectors. The first, is a standalone acoustic-based EP detector as described in section \ref{sec:method-ep-standalone} that provides $P(EOS | x_{1:t} )$ for each feature frame $t$. This model is a 4-layer LSTM network with 128 units, with an output dense layer that performs binary classification for EP decisions at each frame. The second EP detector is an E2E EP detector based on RNNT EOS predictions, with the method described in section \ref{sec:method-ep-e2e}, and the RNNT system described in section \ref{sec:model_details}. An endpoint decision for the baseline system is generated at frame $t$, if  either detectors fire, i.e., $P(EOS | x_{1:t} )> T_{EP}$, or if the EOS symbol is emitted by the RNNT ASR model at frame $t$. The threshold $T_{EP}$ is tuned on the held-out development partition of the transactional queries dataset. The VAD model shares the 4-layer LSTM parameters with the standalone EP, predicts speech/non-speech decision per-frame, and is multi-task trained along with EP. The VAD is used to set an upper-bound latency guardrail of 1740ms based on pause-duration, which is constant for all experiments. The same baseline EP setup is used for experiments with SLURP data, except the end-to-end EP model is derived from the ASR model fine-tuned on SLURP training data.
\vspace{-2mm}
\begin{table*}[t]
  \centering
  \caption{Endpointing results showing relative \% word error rate reductions (WERR), relative \% early endpointing rate reductions (EEPRR), and endpointing latency in msec evaluated on the transactional queries test set. Relative \% early endpoint rate reductions are reported on the partial test set. Negative \% relative values indicate improvements in EEPRR and WER metrics}
  \vspace{-2mm}
  \label{tab:arbitrator-main}
  \begin{tabular}{|l|rrp{1.5cm}|ccc|rr|ccc|}
  \hline
    & \multicolumn{6}{c}{\textbf{Transactional Query test set}}
    & \multicolumn{5}{|c|}{\textbf{Conversational test set}} \\
    \hline
    \multirow{2}{*}{\centering \textbf{Model}} &
    \multicolumn{3}{c|}{\textbf{Accuracy}} & \multicolumn{3}{c|}{\textbf{Latency}} & \multicolumn{2}{c|}{\textbf{Accuracy}} & \multicolumn{3}{c|}{\textbf{Latency}}  \\
    
& \textbf{WERR} & \textbf{EEPRR} & \textbf{EEPRR (partial)} & \textbf{P50} & \textbf{P90} & \textbf{P99} & \textbf{WERR} & \textbf{EEPRR} & \textbf{P50} & \textbf{P90} & \textbf{P99}
\\

    \hline

     Baseline                       & - & -  & \raggedleft - 
     & 300 & 570 & 1110 &  - & - & 570 & 660 & 1140  
    \\

     \hspace{1mm} + Arbitrator
     &  -3.12\% & -16.34\% & \raggedleft -32.45\%  
     & 300 & 600 & 1320 & -0.76\% & -1.48\% & 570 & 660 & 1200
    \\

    \hline
  \end{tabular}
  \vspace{-4mm}
\end{table*}

\begin{table}[t]
\vspace{5mm}
  \centering
  \caption{Endpointing results on SLURP data}
  \vspace{-2mm}
  \label{tab:arbitrator-slurp}
  \begin{tabular}{|l|rr|ccc|}
      \hline
    \multirow{2}{*}{\centering \textbf{Model}} & \multicolumn{2}{c|}{\textbf{Accuracy}} & \multicolumn{3}{c|}{\textbf{Latency}}  \\
& \textbf{WER} & \textbf{EEPR} & \textbf{P50} & \textbf{P90} & \textbf{P99} 
\\
\hline
     Baseline                       & 13.64\% & 2.39\%
     & 270 & 600 & 1320   \\

     \hspace{1mm} + Arbitrator 
     &  13.56\% & 2.19\% 
     & 270 & 600 & 1380   \\
    \hline
\end{tabular}
\vspace{-5mm}
\end{table}

\subsubsection{Experimental EP system: EP Arbitrator} 
\label{sec:exp-setup-arbitrator}
The experimental EP system is a 2nd-pass EP Arbitrator model operating on endpointing candidates generated by the baseline EP system described in Section \ref{subsec:baseline}. The system architecture is as in Fig \ref{fig:arbitrator-detailed} and described in Section \ref{sec:method-ep-arbitrator}. The baseline's 1st-pass EP setup is not modified in our experiments; the 2nd-pass detector is added on top of it.

\noindent \textit{Model Architecture}: The EP Arbitrator model has 4M parameters, and consists of an AcousticEncoder as in Eqn \ref{eqn:acoustic-encoder}, that consumes RNNT encodings $\textbf{h}^{e}_{1:t}$, and uses maxpooling followed by a DNN model to obtain a segment-level acoustic embedding. The TextEncoder consumes the 1-best $\textbf{w}_{t}$ from RNNT at frame $t$, converts it to an embedding sequence, which is passed to a DNN followed by maxpooling to obtain a text embedding. The acoustic and text embeddings are combined with a DNN and softmax as in Eqn \ref{eqn:arbitrator-softmax}. The detector threshold for the arbitrator $T_{arb}$ is tuned on the held-out development partition of the transactional queries dataset.
\\

\noindent \textit{Training}: The EP arbitrator model uses partial recognition from the RNNT model at each frame $t$, to compute model predictions. This is matched with the per-frame EOS labels from forced-alignment to compute cross-entropy loss.\\
The EP arbitrator training data is the same for experiments on all 3 datasets (transactional, conversational and SLURP), and consists only of transactional queries from the voice assistant to validate if the model can work on out-of-domain datasets.
\vspace{-1mm}
\subsection{Metrics}
\vspace{-2mm}
 \noindent \textbf{Word error rate (WER)}: A standard ASR metric defined as the normalized minimum word edit distance between the transcript and predicted recognition. On the SLURP data, we report the absolute WER.

 \noindent \textbf{Word error rate reduction (WERR)}: Relative reductions in word error rate (WER) compared to the baseline system are reported for voice-assistant experiments. Negative values indicate improvements compared to the baseline.
 
 \noindent \textbf{Early endpointing rate (EEPR)}: Early EP rate is the fraction of utterances in a test set where the system endpointed early. For all evaluation sets except the partial test set, an utterance is considered to have an early endpoint when the EP system triggers prior to the ground truth EOS frame. For the transactional queries partial test set, any utterance that has an EP trigger before the end of the audio segment is marked as an early EP, as the partial test set consists of incomplete utterances where the audio after the system cut-off the user is not available for evaluation. On the SLURP data, we report absolute values of EEPR.
 
 \noindent \textbf{Early endpointing rate reduction (EEPRR)}: Relative reductions in early endpointing rate compared to the baseline system, where negative values indicate improvements vs baseline. EEPRR is reported for all experiments on the voice-assistant dataset.\\
\textbf{Endpointing latency}: Percentiles of endpointing latency (P50, P90, P99) are reported in all experiments. The EP latency of an utterance is the time difference between the ground truth EOS and the end of speech frame detected by the EP system, with a minimum frame resolution of 30ms.

\vspace{-2mm}

\section{Results and Discussion}

\subsection{EP Arbitrator can delay endpoints}

In Table \ref{tab:arbitrator-main}, we compare the baseline that uses a 1st-pass-only EP detector (described in section \ref{subsec:baseline}), with an EP system that is augmented with the EP Arbitrator as a 2nd-pass detector (described in section \ref{sec:exp-setup-arbitrator}). The EP arbitrator improves the EEPR by 16.34\% relative and WER by 3.12\% relative on the transactional query test set. On the partial utterances test set, we see a relative improvement of 32.45\% in EEPR. These results show that the EP Arbitrator delays EP decisions for utterances that have longer pauses, hesitations or are semantically incomplete. This is obtained at the cost of no degradation to median latency, a minor increase in P90 latency, and a larger increase in P99. We evaluate the same model, with the same operating point, on conversational data, where we observe a 1.48\% relative improvement in EEPR with no degradation to p90 latency. We consider this an out-of-domain evaluation since both the 1st and 2nd pass endpointing models have not seen any conversational data during training.

In Table \ref{tab:arbitrator-slurp}, we present endpointing results on the SLURP dataset. We observe an 8.37\% relative improvement (2.39\% to 2.19\%) in early EP rate, with no increase in latency at p90, and a small increase in latency at P99, similar to the observations on the voice-assistant datasets. Since EP arbitrator was not trained on SLURP, these results show that our method generalizes to unseen datasets without any explicit training. Further, we report state-of-the-art WER on the SLURP dataset  \footnote{Our reported WER on SLURP (13.64\%) is better than the state-of-the-art results (14.8\%) \cite{sunder2023fine}. The WER without silence padding of utterances is 14.9\%, adding digital silence to the end of each utterance improves this to 13.64\%. This is due to RNNT emission delay, even when the model is trained with delay correction methods \cite{yu21fastemit}, causing deletion errors at the end. This finding is orthogonal to our proposed method but can provide valuable insights to the ASR community.}.
\vspace{-2mm}

\begin{figure}%
    \centering
    \subfloat\centering{{\includegraphics[width=6.5cm]{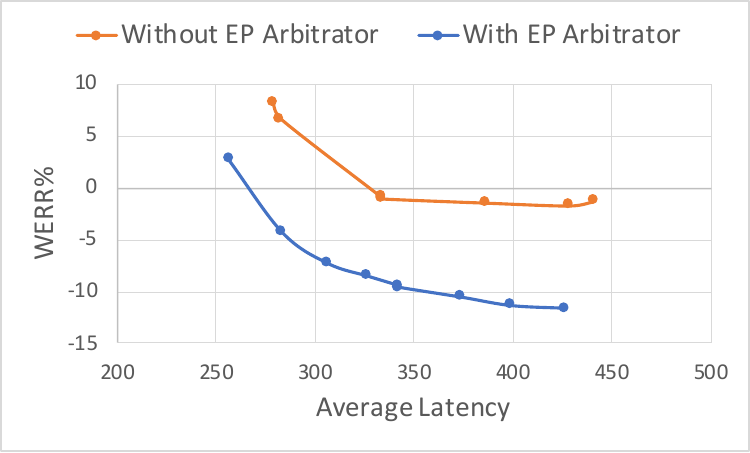} }}%
    \qquad    
    \subfloat\centering{{\includegraphics[width=6.5cm]{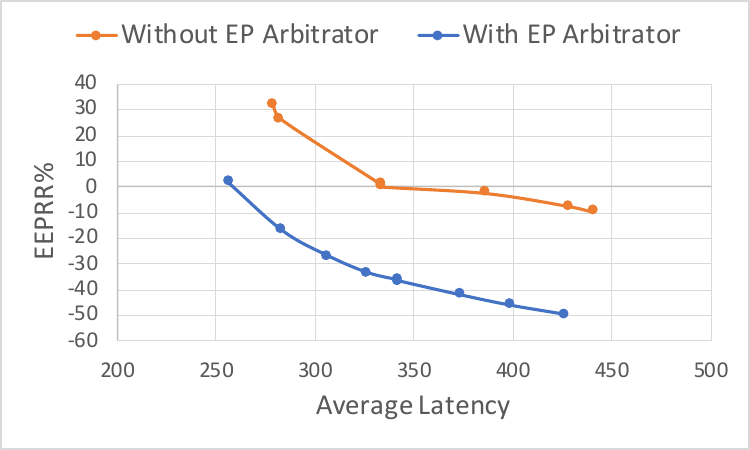} }}%
    \caption{EEPRR\% and WERR\%  vs. Average Latency on transactional queries dataset. WERR, EEPRR are computed with respect to the baseline in Table \ref{tab:arbitrator-main}}%
    \label{fig:pel_vs_eeprwer}%
\end{figure}

\subsection{EP Arbitrator improves early-EP vs latency tradeoff}
\vspace{-1mm}

 In order to verify that the EP Arbitrator performance cannot be obtained by selecting a different operating point on the baseline model, we present the EEPRR vs average latency and WERR vs average latency for both the baseline models and the model including the arbitrator in Figure \ref{fig:pel_vs_eeprwer}, on the transactional query test set. To obtain a baseline curve across various operating points of early EP rate reduction (EEPRR) and latency, we sweep the hyperparameters of the baseline EP system. This includes the first-pass standalone acoustic-EP detector threshold $T_{EP}$, and hyperparameters that control scaling of the RNNT EOS emission probability in decoding \cite{8683109}. For the experimental system, the arbitrator detector threshold $T_{arb}$ is swept to obtain a curve, with the first-pass EP parameters kept unchanged with respect to the baseline.
 
\begin{figure}%
    \centering
    \subfloat\centering{{\includegraphics[width=6.5cm]{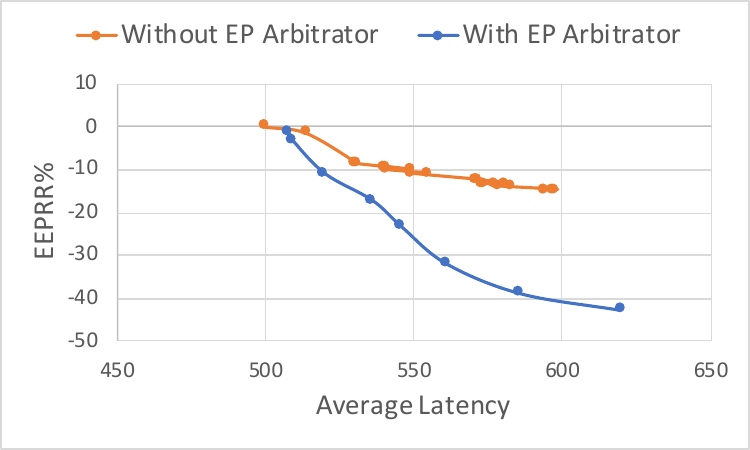} }}%
    \caption{EEPRR\% vs. Avg Latency on Conversational test set, EEPRR is computed with respect to the conversational data baseline in Table \ref{tab:arbitrator-main}}%
    \label{fig:pel_vs_eepr_conv}%
    \vspace{-2mm}
\end{figure}

 The results in Figure \ref{fig:pel_vs_eeprwer} show that the EP arbitrator improves both WER and EEPR at any given average latency operating point compared to the baseline system. Further, in Figure \ref{fig:pel_vs_eepr_conv}, we observe that the same model provides a better EEPR vs latency trade-off on the out-of-domain conversational data, compared to the baseline system, as measured at different average latency operating points (lower on the curves is better, since negative EEPRR indicate improvements). This shows that our method generalizes to out-of-domain conversational data, and improves early EP rate vs latency trade-off. 
 \vspace{-3mm}

\subsection{Analysis with different 1st pass models}
\begin{figure}%
    \centering
    \subfloat\centering{{\includegraphics[width=6.5cm]{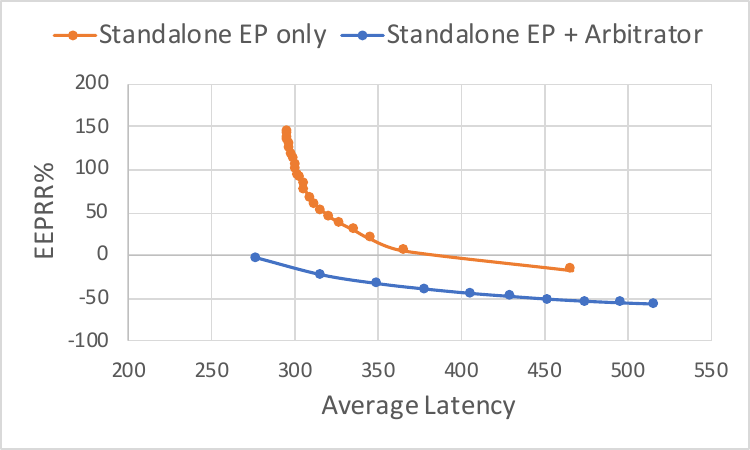} }}%
    \qquad    
    \subfloat\centering{{\includegraphics[width=6.5cm]{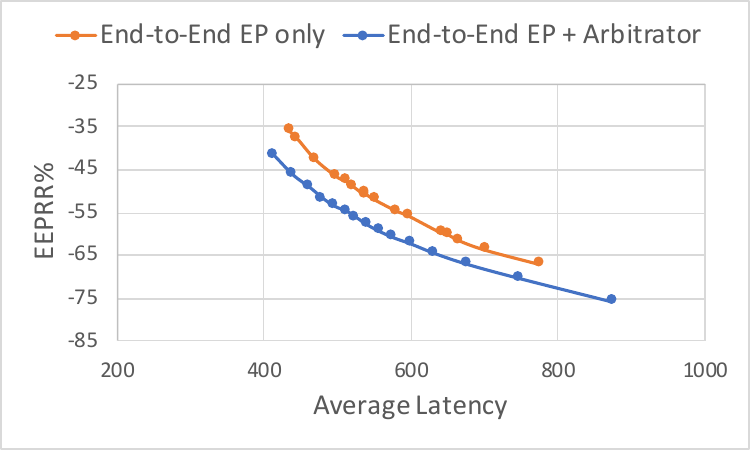}}}%
    \caption{EEPRR\% vs. Average Latency with different first pass models on transactional queries dataset, EEPRR is computed with respect to the baseline in Table \ref{tab:arbitrator-main}}
    \label{fig:ablation_fig}%
\vspace{-3mm}
\end{figure}

To measure efficacy of our method with different 1st-pass EP systems, we present the performance of the EP Arbitrator model on top of two modified baselines that have as 1st-pass EP detector: (1) a standalone acoustic-based EP detector only (2) an end-to-end EP detector only. This is contrast with all the results presented previously where the baseline system included both detectors. Figure \ref{fig:ablation_fig} shows the results of the experiment. With either detector as the 1st-pass, the EP arbitrator model as a 2nd-pass provides a better tradeoff between early EP rate and latency. The EP arbitrator provides higher reduction in EEPR with a stand-alone acoustic-based EP baseline, compared to the E2E EP baseline. This can be explained by the lack of semantic information in the standalone acoustic EP model. EP arbitrator still demonstrates benefits when the end-to-end EP model is used in the first pass. This is owing to the segment level information used by the arbitrator model. 
\section{Conclusions}

Endpoint detection has to trade-off between accuracy and latency, as waiting longer reduces the chance users get cut-off early. We proposed a novel two-pass solution for endpointing to improve the performance of an existing endpointing system, by verifying endpoints with a 2nd-pass EP model termed EP arbitrator. We show improved trade-off between early EP rate and EP latency with our proposed system compared to the baseline. We show that the method generalizes to provide early EP rate improvements on both transactional queries and conversational datasets, without explicit training on conversational data. Further, we propose small modifications to the publicly available SLURP dataset by padding it with digital silence making it suitable for performance evaluation on the endpointing task. Our method improves early EP rate on SLURP by 8\% without degrading latency significantly. Finally, we demonstrate that our method shows improvements regardless of the first-pass EP model.

\bibliographystyle{IEEEtran}
\bibliography{refs}

\end{document}